\documentstyle[12pt,epsfig]{article}
\newcommand{\beq}{\begin{equation}}
\newcommand{\eeq}{\end{equation}}
\newcommand{\beqn}{\begin{eqnarray}}
\newcommand{\eeqn}{\end{eqnarray}}

\newcommand{\stackm}{\stackrel{\scriptstyle <}{{ }_{\sim}}}  
\begin{document}

\thispagestyle{empty}
\def\pubnum{433}
\def\data{December, 1997}
\begin{flushright}
{\parbox{3.5cm}{
UAB-FT-434

December, 1997

hep-ph/9712491
}}
\end{flushright}
\vspace{3cm}
\begin{center}
\begin{large}
\begin{bf}
SUPERSYMMETRIC QUANTUM EFFECTS ON THE HADRONIC WIDTH OF 
A HEAVY CHARGED HIGGS BOSON
IN THE MSSM\\
\end{bf}
\end{large}
\vspace{1cm}
Joan SOL\`A\footnote{Invited talk presented at the {\it International
Workshop on Physics Beyond the Standard Model: from Theory to
Experiment}\,, Valencia, Spain, 
13-17 October, 1997. To appear in the Proceedings.}\\

\vspace{0.25cm} 
Grup de F\'{\i}sica Te\`orica\\ 
and\\ 
Institut de F\'\i sica d'Altes Energies\\ 
\vspace{0.25cm} 
Universitat Aut\`onoma de Barcelona\\
08193 Bellaterra (Barcelona), Catalonia, Spain\\
\end{center}
\vspace{0.3cm}
\hyphenation{super-symme-tric re-nor-ma-li-za-tion}
\hyphenation{com-pe-ti-ti-ve}
\begin{center}
{\bf ABSTRACT}
\end{center}
\begin{quotation}
\noindent
We discuss the QCD and leading electroweak corrections 
to the hadronic width of the charged Higgs boson of the MSSM.
In our renormalization framework, $\tan\beta$ is defined through 
$\Gamma(H^+\rightarrow \tau^+\,\nu_{\tau})$.
We show that a measurement of the hadronic width of $H^\pm$
and/or of the branching ratio of its $\tau$-decay mode  with a modest 
precision of $\sim 20\%$ could be sufficient to unravel the
supersymmetric nature of $H^\pm$ in full consistency with the 
low-energy data from radiative $B$-meson decays.
\end{quotation}
  
\newpage

\baselineskip=6.5mm  %(FOR PREPRINT)
The discovery of a heavy top quark 
at the Tevatron constituted, paradoxically as it may sound,
both a reassuring confirmation
of a long-standing prediction of the Standard Model (SM) of the
electroweak interactions and at the same time the consolidation
of an old and intriguing suspicion, namely,
that the SM cannot be the last word in elementary particle physics.
What are, however, the potential paradigms of new physics at our
disposal?. There are a few good candidates. Notwithstanding, at present  
the only tenable Quantum Field Theory of the strong and the 
electroweak interactions beyond the SM that is able to keep pace with the
SM ability to (consistently) accommodate all known high precision
measurements\,\cite{WdeBoer} is the Minimal Supersymmetric Standard 
Model MSSM\,\cite{MSSM}.
This fact alone,
if we bare in mind the vast amount of high precision data available 
both from low-energy and high-energy physics, 
should justify (we believe) all efforts to search for SUSY in
present day particle accelerators.
Moreover, the MSSM offers  
a starting point for a successful Grand Unified framework\,\cite{Peskin} 
where a radiatively stable low-energy Higgs sector
can survive.

Within the SM the physics of the top quark is intimately connected
with that of the Higgs sector through the Yukawa couplings. One
expects that a first hint of Higgs physics, if ever, should appear in
concomitance with the detailed studies of top quark pheneomenology.
However, if this is true in the SM, the more it should be
in the MSSM where both the Higgs and the top quark sectors are 
virtually ``doubled'' with respect to the SM.
As a consequence, the Yukawa coupling sector is richer in the
supersymmetric model than in the standard one. This
could greatly modify the phenomenology already at the level of
quantum effects on electroweak observables.
As of matter of fact in the MSSM the 
bottom-quark Yukawa coupling may counterbalance the 
smallesness of the bottom
mass at the expense of a large value of $\tan\beta$ --the ratio
$v_2/v_1$ of the vacuum
expectation values of the two Higgs doublets-- the upshot being that
the top-quark and bottom-quark Yukawa couplings in the superpotential
\begin{equation}
h_t={g\,m_t\over \sqrt{2}\,M_W\,\sin{\beta}}\;\;\;\;\;,
\;\;\;\;\; h_b={g\,m_b\over \sqrt{2}\,M_W\,\cos{\beta}}\,,
\label{eq:Yukawas} 
\end{equation}
can be of the same order of magnitude, perhaps even showing up 
in ``inverse'' hierarchy: $h_t<h_b$ for $\tan\beta> m_t/m_b$. 
Notice that due to the perturbative bound
$\tan\beta\stackrel{\scriptstyle <}{{ }_{\sim}}60$ one never reaches
a situation where $h_t<<h_b$.
In a sense $h_t\sim h_b$  could be judged as a natural 
relation in the MSSM.
On the phenomenological side, one should not dismiss the
possibility that the bottom-quark Yukawa coupling could play a
momentous role in the physics of the top quark and of the
Higgs bosons\,\cite{JSola}, to the extend of
drastically changing standard expectations on the observables
associated to them, such as decay widths and 
cross-sections.

It is well-known\,\cite{Hunter} that
the MSSM predicts the existence of two charged Higgs 
pseudoscalar bosons, $H^{\pm}$,
one neutral CP-odd boson, $A^0$, and two neutral
CP-even states, $h^0$ and $H^0$ ($M_{h^0}< M_{H^0}$).
In this talk we wish to emphasize the possibility of
seeing large quantum SUSY signatures in the physics of the MSSM Higgs 
boson decays; specifically,  we shall
concentrate on the potential
supersymmetric quantum effects on the 
decays of the charged Higgs boson of the MSSM as a means to
unveil its hypothetical SUSY nature.

For a heavy charged Higgs boson, the main fermionic decay channel is
the top quark decay mode $H^+\rightarrow t\,\bar{b}$, whose partial 
width is essentially the full hadronic width of $H^+$.
In fact, the other two standard fermionic decay modes are 
$H^+\rightarrow c\,\bar{s}$ and $H^+\rightarrow \tau^+\,\nu_{\tau}$.
However, for any charged Higgs mass $M_H$ and $\tan\beta>2$, 
the branching ratio of the former decay is  negligibly small, whereas
that of the latter is subdominant but not negligible (see later on).
The bare interaction Lagrangian
describing the $H^{+}\,t\,\bar{b}$-vertex in the MSSM reads
as follows\,\cite{Hunter}: 
\beq
{\cal L}_{Htb}={g\,V_{tb}\over\sqrt{2}M_W}\,H^+\,\bar{t}\, [m_t\cot\beta\,P_L
+ m_b\tan\beta\,P_R]\,b+{\rm h.c.}\,,
\label{eq:LtbH}
\eeq
where $P_{L,R}=1/2(1\mp\gamma_5)$ are the chiral projector operators
and $V_{tb}$ is the CKM
matrix element -- henceforth we set $V_{tb}=1$.
The corresponding counterterm Lagrangian 
follows right away after
re-expressing everything in terms of renormalized parameters and 
fields in the standard electroweak 
on-shell scheme\,\cite{BSH}. It takes on the form:
\beq
\delta{\cal L}_{Htb}={g\over\sqrt{2}\,M_W}\,H^+\,\bar{t}\left[
\delta C_L\ m_t\,\cot\beta\,\,P_L+
\delta C_R\ m_b\,\tan\beta\,P_R\right]\,b
+{\rm h.c.}\,,
\label{eq:LtbH2}
\eeq
with
\beqn
\delta C_L &=& {\delta m_t\over m_t}-{\delta v\over v}
+\frac{1}{2}\,\delta Z_{H^+}+\frac{1}{2}\,\delta Z_L^b+\frac{1}{2}
\,\delta Z_R^t
-\delta\tan\beta+\delta Z_{HW}\,\tan\beta\,,\nonumber\\
\delta C_R &=& {\delta m_b\over m_b}-{\delta v\over v}
+\frac{1}{2}\,\delta Z_{H^+}+\frac{1}{2}\,\delta Z_L^t+\frac{1}{2}
\,\delta Z_R^b
+\delta\tan\beta-\delta Z_{HW}\,\cot\beta\,.\nonumber\\
\eeqn 
Here $\delta v$ is the counterterm 
for $v=\sqrt{v_1^2+v_2^2}=\sqrt{2}\,M_W/g$;
$\delta Z_{H}$ and $\delta Z_{HW}$ stand
respectively for the charged Higgs and mixed
$H-W$ wave-function renormalization factors. The remaining 
are standard wave-function and mass renormalization counterterms
for the fermion external lines\,\cite{BSH}.

We remark the counterterm $\delta\tan\beta$, 
which is fixed here by the renormalization condition
that the parameter
$\tan\beta$ is inputed from the
tree-level expression for
$\Gamma (H^+\rightarrow \tau^+\,\nu_{\tau})$
either in the $\alpha$ or in the $G_F$ schemes\,\footnote{One could
also define $\tan\beta$ through the $\tau$-decay of the
CP-odd Higgs boson, $A^0\rightarrow \tau^+\,\tau^-$,
and then compute quantum corrections to 
$\Gamma (H^+\rightarrow \tau^+\,\nu_{\tau})$\,\cite{MAD}.
Conversely, in the framework of the present paper we could
compute one-loop effects on the partial width of 
$A^0\rightarrow \tau^+\,\tau^-$.}.
From this it follows that\,\cite{CGGJS} 
\beq
{\delta\tan\beta\over \tan\beta}
=\frac{1}{2}\left(
\frac{\delta M_W^2}{M_W^2}-\frac{\delta g^2}{g^2}\right)
-\frac{1}{2}\delta Z_H
+\cot\beta\, \delta Z_{HW}+ 
\Delta_{\tau}\,,
\label{eq:deltabeta}
\eeq    
and so the various one-loop diagrams
for $H^+\rightarrow t\,\bar{b}$ 
can be parametrized in terms of two form factors $F_L$,
$F_R$ and the remaining
counterterms:
\beq
i\,\Lambda = {i\,g\over\sqrt{2}\,M_W}
\,\left[m_t\,\cot\beta\,(1+\Lambda_L)\,P_L
 + m_b\,\tan\beta\,(1+\Lambda_R)\,P_R\right]\,,
\label{eq:AtbH}
\eeq
where
\beqn
\Lambda_L & = & F_L+{\delta m_t\over m_t}
+\frac{1}{2}\,\delta Z_L^b+\frac{1}{2}\,\delta Z_R^t-\Delta_{\tau}\nonumber\\
& - & {\delta v^2\over v^2}+\delta Z_{H^+}+(\tan\beta-\cot\beta)\,\delta Z_{HW}
 \,,\nonumber\\
\Lambda_R &=& F_R+{\delta m_b\over m_b}
+\frac{1}{2}\,\delta Z_L^t+\frac{1}{2}\,\delta Z_R^b
+\Delta_{\tau}\,.
\label{eq:lambdaLR}
\eeqn
In these equations, $\Delta_{\tau}$ involves the complete set of
MSSM one-loop effects on the $\tau$-lepton decay width of $H^\pm$.

%%%%%%%%%%%%%%%%%%%%%%%%%%%%%%%%%%%%%%%%%%%%%%%%%%%%%%%%%%%%%%%%%%%%%%%
%Fig.1
\begin{figure}
\centering
\mbox{\epsfig{file=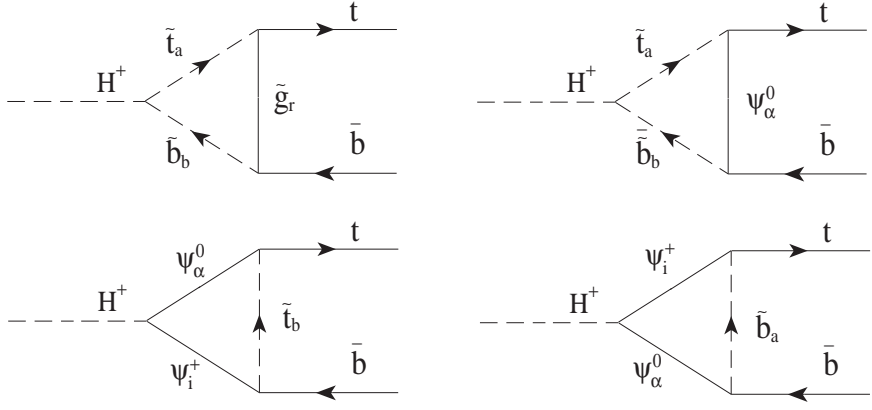,width=11.5cm}}
%%%
%%%%%%  Figures Centrades Verticalment
%%%
%\noindent
\caption
{SUSY-QCD and SUSY-EW one-loop vertices for 
$H^+\rightarrow t\,\bar{b}$.}
%%\end{center}
\end{figure}
%%%%%%%%%%%%%%%%%%%%%%%%%%%%%%%%%%%%%%%%%%%%%%%%%%%%%%%%%%%%%%%%%%%%%%%%%%
%%%%%%%%%%%%%%%%%%%%%%%%%%%%%%%%%%%%%%%%%%%%%%%%%%%%%%%%%%%%%%%%%%%%%%%
%Fig.2 
\begin{figure}
\centering 
\mbox{\epsfig{file=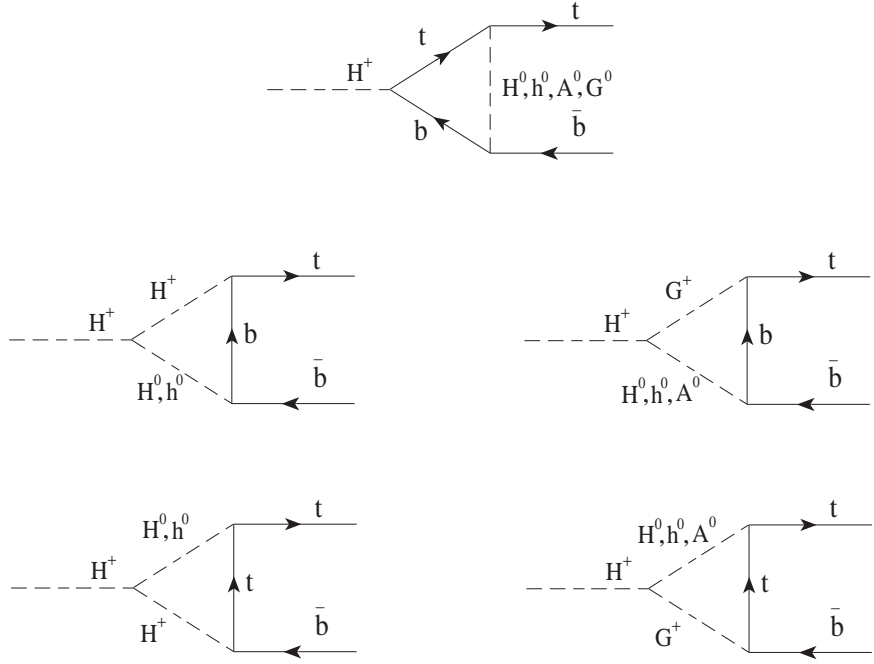,width=11.5cm}}
%%%
%%%%%%  Figures Centrades Verticalment
%%%
%\noindent
\caption
{One-loop vertices from the Higgs sector of the MSSM for 
$H^+\rightarrow t\,\bar{b}$.}
%%\end{center}
\end{figure}
%%%%%%%%%%%%%%%%%%%%%%%%%%%%%%%%%%%%%%%%%%%%%%%%%%%%%%%%%%%%%%%%%%%%%%%%%%%%
The basic free parameters of our analysis, in the electroweak sector, are 
contained in the
stop and sbottom mass matrices: 
\begin{equation}
{\cal M}_{\tilde{t}}^2 =\left(\begin{array}{cc}
M_{\tilde{t}_L}^2+m_t^2+\cos{2\beta}({1\over 2}-
{2\over 3}\,s_W^2)\,M_Z^2 
 &  m_t\, M_{LR}^t\\
m_t\, M_{LR}^t &
M_{\tilde{t}_R}^2+m_t^2+{2\over 3}\,\cos{2\beta}\,s_W^2\,M_Z^2\,  
\end{array} \right)\,
\label{eq:stopmatrix}
\end{equation}
\begin{equation}
{\cal M}_{\tilde{b}}^2 =\left(\begin{array}{cc}
M_{\tilde{b}_L}^2+m_b^2+\cos{2\beta}(-{1\over 2}+
{1\over 3}\,s_W^2)\,M_Z^2 
 &  m_b\, M_{LR}^b\\
m_b\, M_{LR}^b &
M_{\tilde{b}_R}^2+m_b^2-{1\over 3}\,\cos{2\beta}\,s_W^2\,M_Z^2\,  
\end{array} \right)\,
\label{eq:sbottommatrix}
\end{equation}
with 
\beq
M_{LR}^t=A_t-\mu\cot\beta\,, \ \ \ \ M_{LR}^b=A_b-\mu\tan\beta\,, 
\eeq
$\mu$ being the SUSY Higgs mass parameter in the superpotential.
The $A_{t,b}$ are the trilinear soft SUSY-breaking parameters and the
$M_{{\tilde{q}}_{L,R}}$ are soft SUSY-breaking masses.
By $SU(2)_L$-gauge invariance we must have $M_{\tilde{t}_L}=M_{\tilde{b}_L}$,
whereas $M_{{\tilde{t}}_R}$, $M_{{\tilde{b}}_R}$ are in general independent
parameters. We denote by $m_{\tilde{t}_1}$ and $m_{\tilde{b}_1}$ the
lightest stop and sbottom mass eigenvalues.
In the strong supersymmetric gaugino
sector, the basic parameter is the
gluino mass, $m_{\tilde{g}}$.

The one-loop vertices contributing to the above 
on-shell form factors in the MSSM are displayed
in Figs.\,1 and 2.
%% 
%%%%%%%%%%%%%%%%%%%%%%%%%%%%%%%%%%%%%%%%%%%%%%%%%%%%%%%%%%%%%%%%%%%%%%%
%Fig.3
\begin{figure}
\centering
\mbox{\epsfig{file=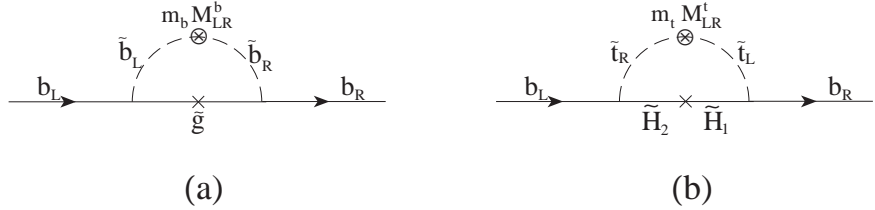,width=11.5cm}}
%%%
%%%%%%  Figures Centrades Verticalment
%%%
%\noindent
\caption
{Leading SUSY-QCD (a) and SUSY-EW (b)
contributions to $\delta m_b/m_b$ in the
electroweak-eigenstate basis. The $\tilde{H}_i\, (i=1,2)$ are the
charged higgsinos.}
%%\end{center}
\end{figure}
%%%%%%%%%%%%%%%%%%%%%%%%%%%%%%%%%%%%%%%%%%%%%%%%%%%%%%%%%%%%%%%%%%%%%%%%%%%% 
Specifically, in Fig.\,1 we show the SUSY-QCD 
contributions from gluinos $\tilde{g}_r\, (r=1,...,8)$ and 
bottom- and top-squarks $\tilde{b}_a,\tilde{t}_b\, (a,b=1,2)$. 
Also shown are
the leading supersymmetric  electroweak effects (SUSY-EW) driven by 
the Yukawa couplings (\ref{eq:Yukawas}); the latter effects consist of
the genuine (R-odd) SUSY contributions from
chargino-neutralinos $\Psi^+_{i},\, \Psi^0_{\alpha}
\,(i=1,2; \alpha=1,..,4)$
and bottom- and top-squarks.
On the other hand in Fig.\,2 we detail
the various Higgs and Goldstone
boson graphs -- which we compute in the Feynman gauge.

As we said, the three-point functions in Figs.\,1 and 2 are 
renormalized Green functions,
so that appropriate counterterms for all the fermionic and Higgs
wave-function external lines are
already included. In particular, in Fig.\,3 we single out the
(finite) leading parts of the bottom quark
supersymmetric self-energy loops. They
feed the form factor $\Lambda_R$ in eq.(\ref{eq:lambdaLR})
through the bottom mass counterterm
and are numerically very relevant:
\beqn
\left({\delta m_b\over m_b}\right) &\simeq&
-{2\alpha_s(m_t)\over 3\pi}\,m_{\tilde{g}}\,\mu\tan\beta\,
I(m_{\tilde{b}_1},m_{\tilde{b}_2},m_{\tilde{g}})\nonumber\\
&-&{h_t^2\over 16\pi^2}\,\mu\tan\beta\,A_t\,
I(m_{\tilde{t}_1},m_{\tilde{t}_2},\mu)\,,
\label{eq:dmbSQCEW}
\eeqn
where
\beq
I(m_1,m_2,m_3)=
{m_1^2\,m_2^2\ln{m_1^2\over m_2^2}
+m_2^2\,m_3^2\ln{m_2^2\over m_3^2}+m_1^2\,m_3^2\ln{m_3^2\over m_1^2}\over
 (m_1^2-m_2^2)\,(m_2^2-m_3^2)\,(m_1^2-m_3^2)}\,.
\label{eq:I123}
\eeq
%%%%%%%%%%%%%%%%%%%%%%%%%%%%%%%%%%%%%%%%%%%%%%%%%%%%%%%%%%%%%%%%%%%%%%%
%Fig.4
\begin{figure}[h]
\centering
\mbox{\epsfig{file=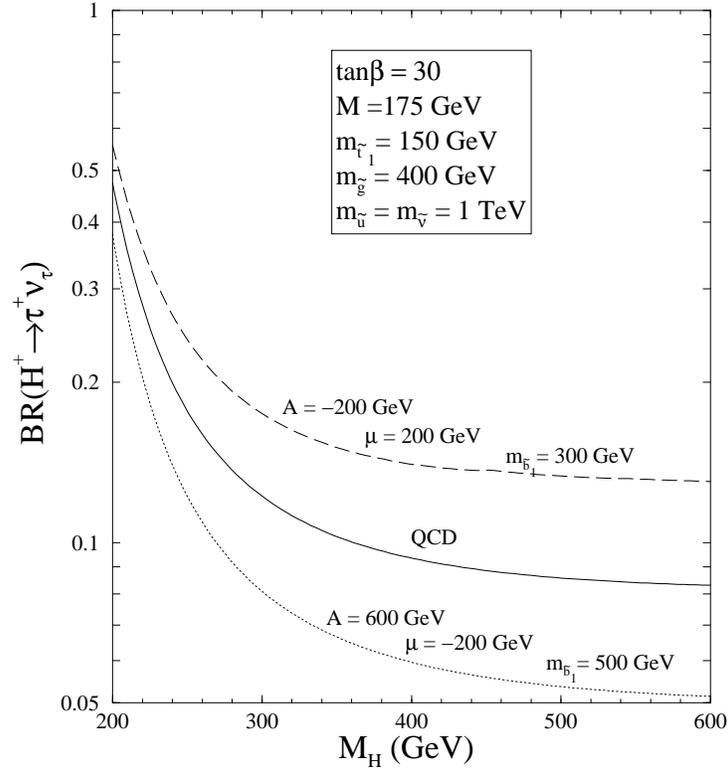,height=10.5cm}}
%%%
%%%%%%  Figures Centrades Verticalment
%%%
%\noindent
\caption
{The branching ratio of $H^+\rightarrow \tau^+\,\nu_{\tau}$ as a
function of the charged Higgs mass for fixed values of the other
parameters; $A$ is a common value for the trilinear couplings.
The central curve includes the standard QCD effects only.}
%%\end{center}
\end{figure}
%%%%%%%%%%%%%%%%%%%%%%%%%%%%%%%%%%%%%%%%%%%%%%%%%%%%%%%%%%%%%%%%%%%%%%%%%%%%
%%%%%%%%%%%%%%%%%%%%%%%%%
For the numerical analysis we wish to single out the Tevatron accessible
window for the charged Higgs mass 
\beq
m_t\stackm M_H\stackm 300\,GeV\,.
\label{eq:interval}
\eeq 
This window is especially
significant in that the CLEO measurements\,\cite{CLEO} of
$BR(b\rightarrow s\,\gamma)$ forbid most of this domain
within the context of a generic two-Higgs-doublet model ($2$HDM).
In contrast, within the MSSM the mass interval
(\ref{eq:interval}) is perfectly consistent\,\cite{CGGJS2} with the 
$b\rightarrow s\,\gamma$-allowed region.
At this point it may be wise to make the following remark. Although the 
inclusion of the NLO effects on the charged Higgs corrected amplitude 
may considerably shift\,\cite{Ciucini} the range (\ref{eq:interval})
up to higher values of $M_H$, within the strict $2$HDM,
the NLO corrections on the SUSY amplitudes 
have {\it not} been computed, and so as in the LO case they could also
contribute to compensate the Higgs counterpart.  
We recall that for light charginos and stops ($\stackm 100\,GeV$) the
CLEO data\,\cite{CLEO} on $b\rightarrow s\,\gamma$ allow\,\cite{Ng}
supersymmetric charged Higgs bosons to exist in the
kinematical window enabling the top quark decay
$t\rightarrow H^+\,b$\,\cite{Guasch4}, which is crossed to 
the one under consideration.

In the following we present some of the results of the numerical
analysis\,\cite{CGGJS2}. It was already shown\,\cite{Ricard} that SUSY-QCD
effects decrease significantly
with increasing sbottom masses. Even so, for $m_{\tilde{b}_1}$ of a few
hundred GeV (e.g. $100-200\,GeV$) they remain important. 
Here we wish to concentrate on a region of the MSSM
parameter space where the only relevant charged Higgs boson decays
are $H^+\rightarrow t\,\bar{b}$ and $H^+\rightarrow \tau^+\,\nu_{\tau}$.
An scenario like this is possible if the squarks are sufficiently heavy that
the direct SUSY Higgs decays into top and sbottom squarks, namely
$H^{+}\rightarrow\tilde{t}_i\,\bar{\tilde{b}}_j$\,\cite{BartlMaj},
are not possible. Moreover, the $H^+\rightarrow W^+\,h^0$ decay which 
is sizeable enough at low $\tan\beta$ it becomes extremely
depleted\,\cite{Ricard} at high $\tan\beta$. Finally, the decays into 
charginos and neutralinos, $H^+\rightarrow \chi^+_i\,\chi_{\alpha}^0$,
are not $\tan\beta$-enhanced and remain negligible. 
In practice we may effectively set up the desired scenario
if we assume that the sbottoms are rather heavy 
(i.e. typically $m_{\tilde{b}_1}\geq 300\,GeV$). 
It is known\,\cite{CGGJS2} that this
assumption is compatible with the MSSM analysis of
$b\rightarrow s\, \gamma$ at large $\tan\beta$. Therefore, we expect
that the SUSY-QCD effects will be somewhat depressed, but now the question
is whether this withdrawal of the gluino-squark corrections  
could be compensated by the
SUSY-EW contributions triggered by the Yukawa couplings.
%%%%%%%%%%%%%%%%%%%%%%%%%%%%%%%%%%%%%%%%%%%%%%%%%%%%%%%%%%%%%%%%%%%%%%
%Fig.5
\begin{figure}
\centering
        \begin{tabular}{cc}
          \mbox{\epsfig{file=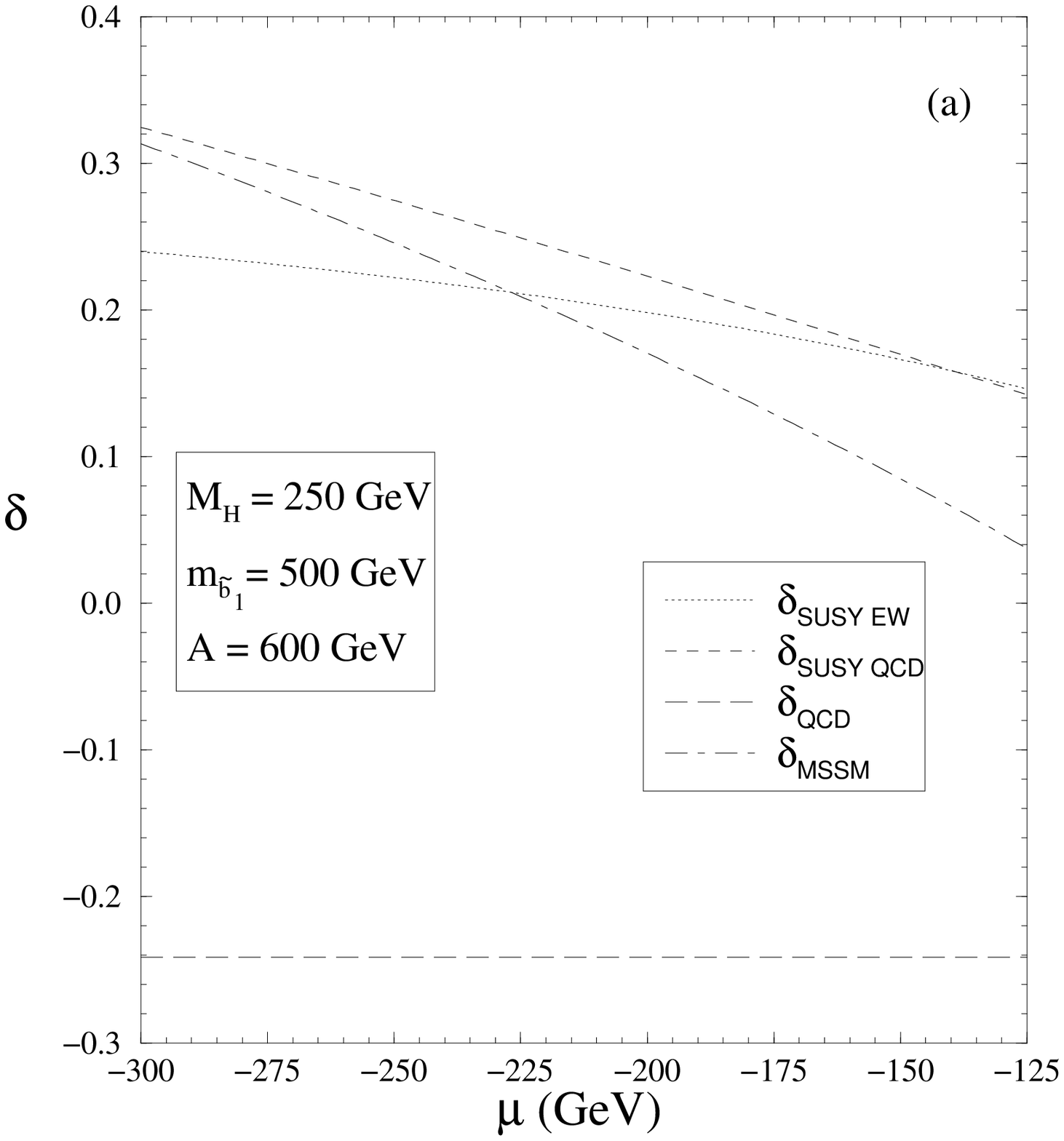,width=5.5cm}}&
          \mbox{\epsfig{file=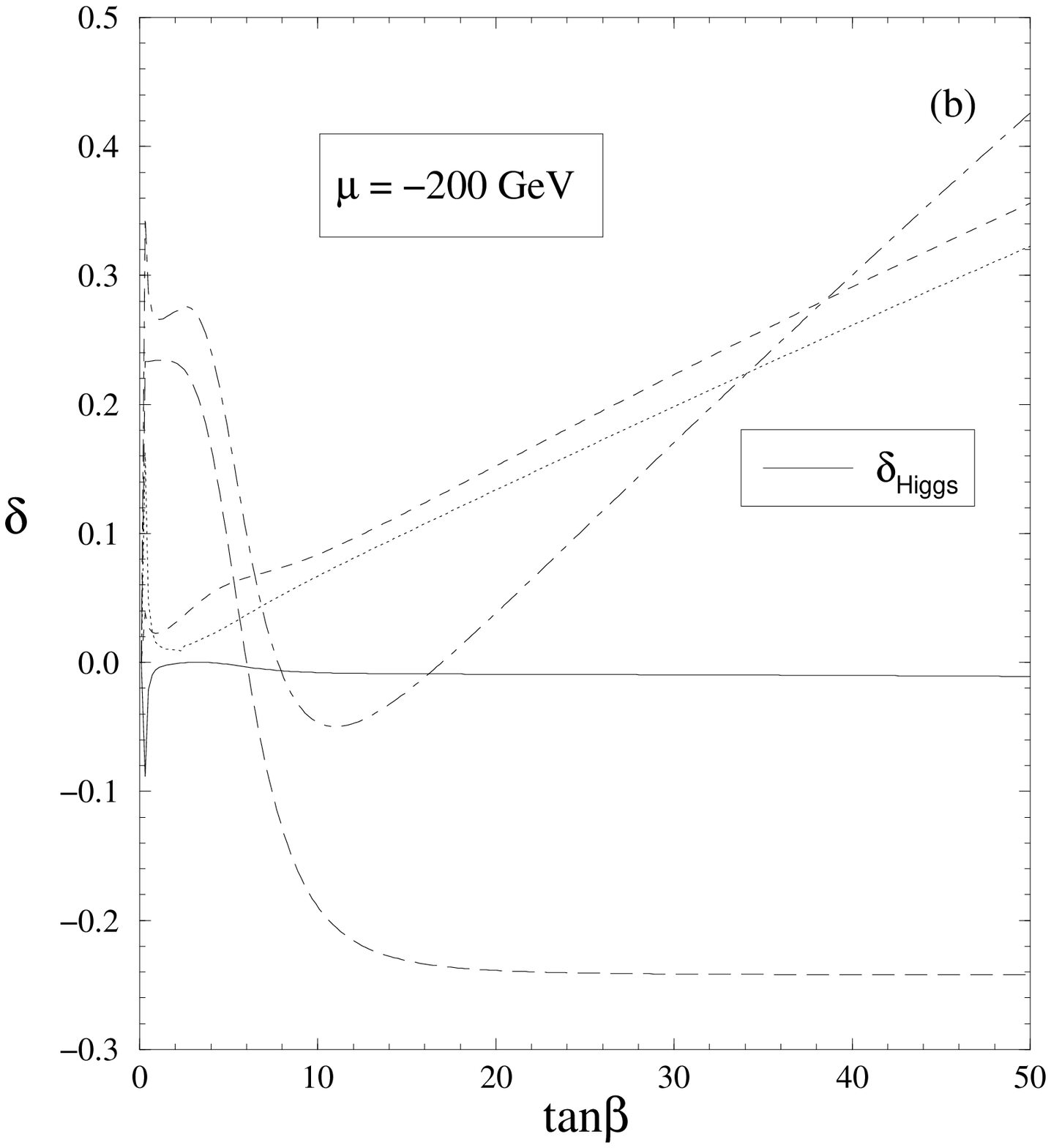,width=5.5cm}} \\[0.5cm]
          \mbox{\epsfig{file=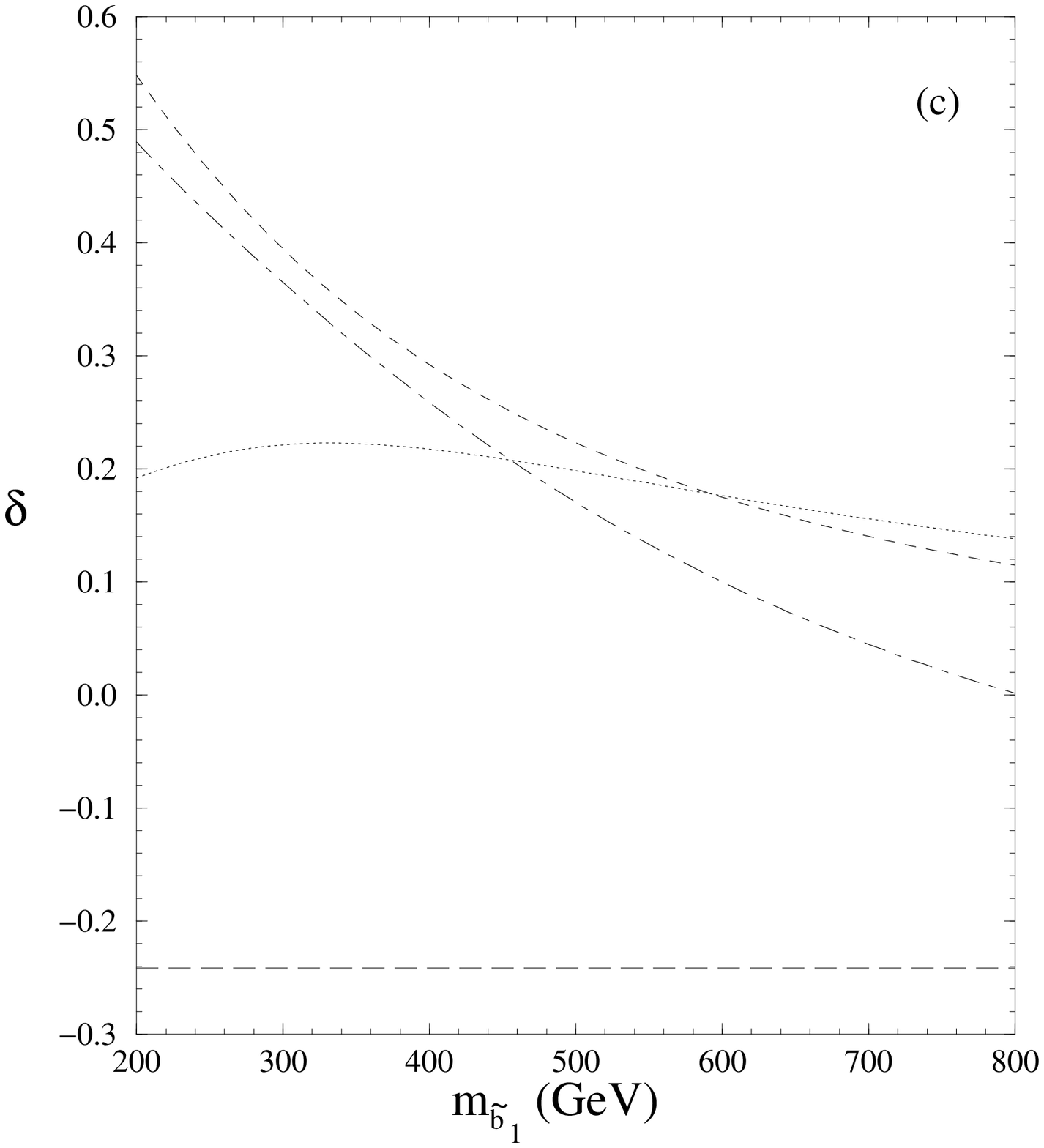,width=5.5cm}}&
          \mbox{\epsfig{file=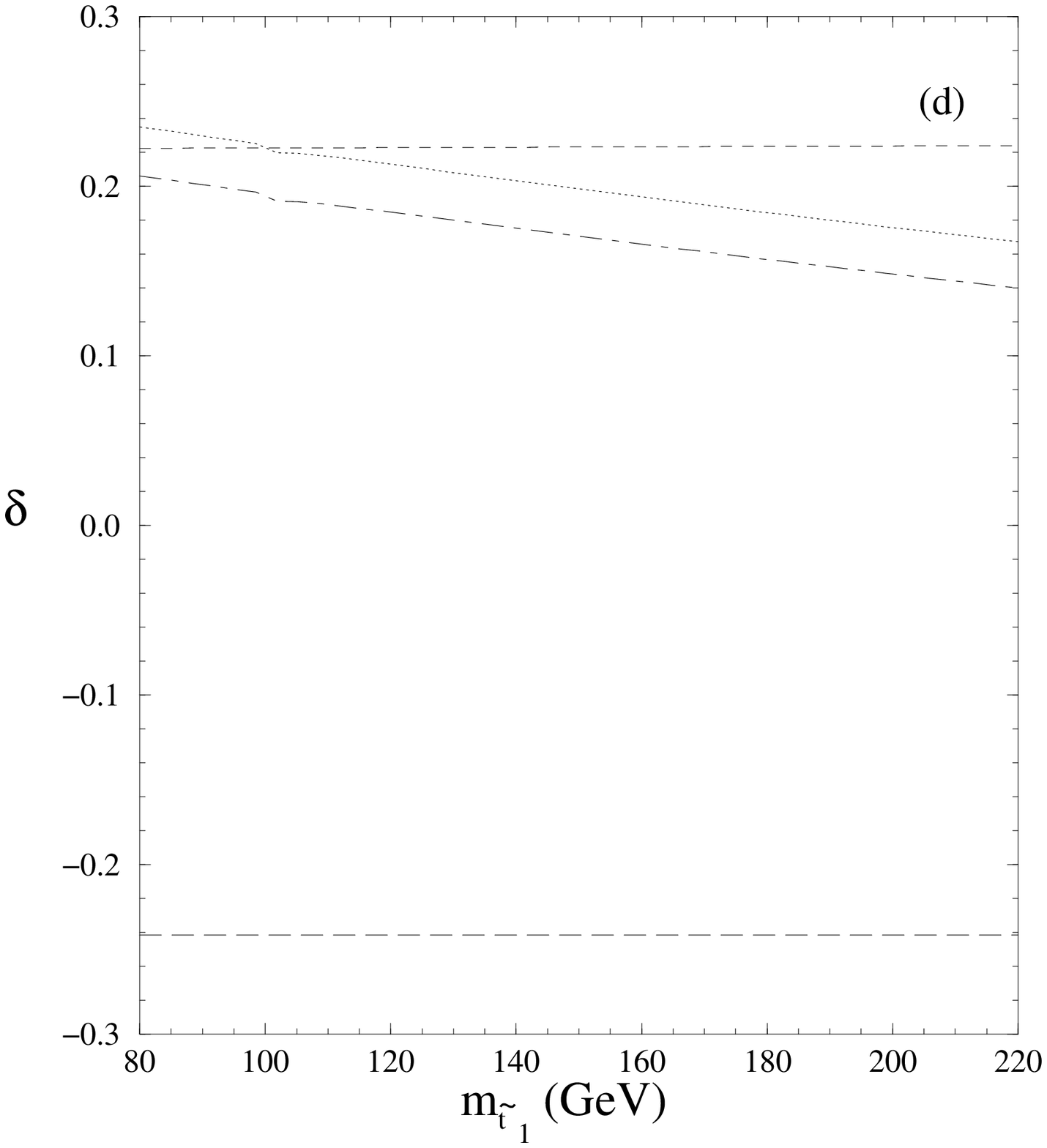,width=5.5cm}} 
        \end{tabular}
%%%
%%%%%%  Figures Centrades Verticalment
%%%
%\noindent
\caption
{{\bf (a)} The SUSY-EW, SUSY-QCD, standard QCD
and full MSSM contributions, eq.(\protect{\ref{eq:deltag}}),
as a function of
$\mu$; {\bf (b)} As in (a), but as a function of
$\tan\beta$. Also shown in (b) is the Higgs contribution,
$\delta_{\rm Higgs}$;
{\bf (c)} As in (a), but as a function of
$m_{\tilde{b}_1}$; {\bf (d)} As a function of
$m_{\tilde{t}_1}$. Remaining inputs as in Fig.\,4.}
%%\end{center}
\end{figure}
%%%%%%%%%%%%%%%%%%%%%%%%%%%%%%%%%%%%%%%%%%%%%%%%%%%

In Fig.\,4 we confirm that $BR (H^+\rightarrow \tau^+\,\nu_{\tau})$
at high $\tan\beta$ is quite large in the Higgs mass interval
(\ref{eq:interval}) and that it never decreases below $5-10\%$ in
the whole mass range up to about $1\,TeV$. Hence this process can
safely be used to input $\tan\beta$ from experiment and in this way 
we are ready to study the
evolution of the quantum 
corrections to the original decay $H^+\rightarrow t\,\bar{b}$
as a function of the most significant parameters. These are shown in
Figs.\,5a-5d. 
The corrections are defined in terms of the quantity
\beq
\delta={\Gamma (H^+\rightarrow t\,\bar{b})-
\Gamma_0 (H^+\rightarrow t\,\bar{b})\over
\Gamma_0 (H^+\rightarrow t\,\bar{b})}\,,
\label{eq:deltag}
\end{equation}
which traces the size of the effect with respect to the 
tree-level width. 
The MSSM correction (\ref{eq:deltag}) includes the full QCD
yield (both from gluons\,\cite{Oakes} and gluinos\,\cite{Ricard})
at ${\cal O}(\alpha_s)$ plus all the 
leading MSSM electroweak effects\,\cite{CGGJS2} driven by the
Yukawa couplings (\ref{eq:Yukawas}). 

In order to assess the impact of the electroweak
effects, we demonstrate that a  typical set  of inputs
can be chosen such that the SUSY-QCD and 
SUSY-EW outputs are of comparable size.
In Figs.\,5a and 5b  we display
$\delta$, eq.(\ref{eq:deltag}), as a function respectively of $\mu<0$ 
and $\tan\beta$ for fixed values of the other parameters (within the 
$b\rightarrow s\,\gamma$ allowed region). Remarkably, in spite of the fact  
that all sparticle masses are beyond the scope of LEP$\,200$ the 
corrections are fairly large. 
We have individually plot the SUSY-EW, SUSY-QCD, standard QCD
and total MSSM effects. 
The Higgs-Goldstone boson corrections
are isolated only in Fig.\,5b just to make clear that
they add up non-trivially to a very
tiny value in the whole large $\tan\beta$ range, and that 
only in the small corner $\tan\beta<1$  they can be significant.
We point out that for general (non-SUSY) $2$HDM's,
the Higgs sector
may give a more relevant contribution, whose origin
stems not only from
the top quark Yukawas\,\cite{Yang}, but also from the bottom
quark sector\,\cite{HtbCoarasa2}.

In Figs.\,5c-5d we render the various corrections (\ref{eq:deltag})
as a function of the relevant squark masses.
For $m_{\tilde{b}_1}<200\,GeV$ we observe 
(Cf. Fig.\,5c) that the SUSY-EW contribution is non-negligible 
($\delta_{SUSY-EW}\simeq +20\%$) but 
the SUSY-QCD loops induced by squarks and gluinos are by far the
leading SUSY effects ($\delta_{SUSY-QCD}> 50\%$) -- the standard QCD
correction staying invariable over $-20\%$ and the standard EW correction
(not shown) being negligible. 
In contrast, for larger and larger $m_{\tilde{b}_1}>300\,GeV$, say
$m_{\tilde{b}_1}=400$ or $500\,GeV$, and fixed 
stop mass at a moderate value $m_{\tilde{t}_1}=150\,GeV$, the
SUSY-EW output is longly sustained whereas the SUSY-QCD one 
steadily goes down.
However, the total SUSY pay-off adds up to about
$+40\%$ and the net MSSM yield still
reaches a level around $+20\%$, i.e. in this case being of equal
value but opposite in sign to the conventional QCD result.
A qualitatively
distinct and quantitatively sizeable quantum
signature like that could not be missed!.

From another point of view, the virtual SUSY effects on
$\Gamma(H^+\rightarrow t\,\bar{b})$ could also be pinned down indirectly
from the renormalization effect on the branching ratio of the $\tau$-mode
$H^+\rightarrow\tau^+\,\nu_{\tau}$.
Indeed, returning to Fig.\,4
and taking the standard QCD-corrected branching ratio
of the $\tau$-lepton decay mode of $H^+$ (central curve in that figure)
as a fiducial quantity, we see that
$BR(H^+\rightarrow\tau^+\,\nu_{\tau})$ undergoes an effective 
MSSM renormalization of order\, $\pm (40-50)\%$. 
The sign of this correction is given by the
sign of $\mu$. In practice this could be a good alternative
method to experimentally tag this effect, for the observable
$BR(H^+\rightarrow\tau^+\,\nu_{\tau})$ should
be directly measurable from the cross-section for 
$\tau$-production at e.g. the Tevatron and the LHC. 

In summary, the SUSY contributions to the hadronic 
width  $\Gamma(H^+\rightarrow t\,\bar{b})$ could be quite large,
namely of the order of several ten percent. These results have
been obtained within the domain of experimental 
compatibility of the MSSM with $b\rightarrow s\,\gamma$.
In general the leading supersymmetric contribution stems from the
SUSY-QCD sector. However, we have produced an scenario where the
SUSY-EW could be equally important.
The upshot is that the whole range of charged
Higgs masses up to about $1\,TeV$ could be probed and, within the
present renormalization framework, its potential supersymmetric nature be 
unravelled through a measurement of $\Gamma (H^+\rightarrow t\,\bar{b})$ 
with a modest precision of $\sim 20\%$. Alternatively, one could look
for indirect SUSY quantum effects on
the branching ratio of $H^+\rightarrow \tau^+\,\nu_{\tau}$ 
by measuring this observable to within
a similar degree of precision. 
At the end of the day we have found that the physics of the
combined decays
$H^+\rightarrow t\,\bar{b}$ 
and $H^+\rightarrow \tau^+\,\nu_{\tau}$  could be the ideal
environment where to target our 
search for large non-SM quantum effects that would
strongly hint at the SUSY nature of the charged Higgs.

%%%%%%%%%%%%%%%%%%%%%%%%%%%%%%%%%%%%%%%%%%%%%%%%%%%%%%%%%%%%%%%%%%%%%%%%%%%%%%
\vspace{1cm}
{\bf Acknowledgements}:
This work has been partially supported by CICYT 
under project No. AEN93-0474. I am grateful to Toni Coarasa,
David Garcia, Jaume Guasch and R.A. Jim\'enez for a fruitful 
collaboration. I am also indebted to Toni for helping me to
prepare the plots for the writing up of my talk.
\noindent

%%%%%%%%%%%%%%%%%%%%%%%%%%%%%%%%%%%%%%%%%%%%%%%%%%%%%%%%%%%%%%%%%%%

\end{document}